\def\papertitle{A DDSP Framework for Adaptive Room Equalization}
\def\paperauthorA{Fernando Marcos-Macías}
\def\paperauthorB{María Pilar Daza-Llin}
\def\paperauthorC{Mateo Cámara}
\def\paperauthorD{José Luis Blanco}

\documentclass[twoside,a4paper]{article}
\usepackage{etoolbox}

\usepackage[print]{dafx26v3}

\usepackage{amsmath,amssymb,amsfonts,amsthm}
\usepackage{siunitx}
\usepackage{euscript}
\usepackage[T1]{fontenc}
\usepackage{multirow}
\usepackage[utf8]{inputenc}
\usepackage{ifpdf}
\usepackage[english]{babel}
\usepackage{caption}
\usepackage{subfig} 
\usepackage{color}
\usepackage{booktabs}
\usepackage{lipsum}

\usepackage{soul}

\usepackage{float}

\input glyphtounicode
\ninept

\newcounter{numauth}
\newcounter{listcnt}
\newcommand\authcnt[1]{\ifdefined#1 \stepcounter{numauth} \fi}

\newcommand\addauth[1]{
\ifdefined#1 
\stepcounter{listcnt}
\ifnum \value{listcnt}<\value{numauth}
\appto\authorslist{, #1}
\else
\appto\authorslist{~and~#1}
\fi
\fi}
\authcnt{\paperauthorB}
\authcnt{\paperauthorC}
\authcnt{\paperauthorD}
\authcnt{\paperauthorE}
\authcnt{\paperauthorF}
\authcnt{\paperauthorG}
\authcnt{\paperauthorH}
\authcnt{\paperauthorI}
\authcnt{\paperauthorJ}
\def\authorslist{\paperauthorA}
\addauth{\paperauthorB}
\addauth{\paperauthorC}
\addauth{\paperauthorD}
\addauth{\paperauthorE}
\addauth{\paperauthorF}
\addauth{\paperauthorG}
\addauth{\paperauthorH}
\addauth{\paperauthorI}
\addauth{\paperauthorJ}

\usepackage{times}

\newif\ifpdf
\ifx\pdfoutput\relax
\else
   \ifcase\pdfoutput
      \pdffalse
   \else
      \pdftrue
   \fi
\fi

\ifpdf 
  \usepackage[pdftex,
    pdftitle={\papertitle},
    pdfauthor={\authorslist},
    pdfsubject={Proceedings of the 29th International Conference on Digital Audio Effects (DAFx26)},
    colorlinks=false, 
    bookmarksnumbered, 
    pdfstartview=XYZ 
  ]{hyperref}
  \usepackage[pdftex]{graphicx}
\else 
  \usepackage[dvips]{epsfig,graphicx}
  \usepackage[dvips,
    pdftitle={\papertitle},
    pdfauthor={\authorslist},
    pdfsubject={Proceedings of the 29th International Conference on Digital Audio Effects (DAFx26)},
    colorlinks=false, 
    bookmarksnumbered, 
    pdfstartview=XYZ 
  ]{hyperref}
\fi
\usepackage[hypcap=true]{caption}
\title{\papertitle}

\affiliation
{\paperauthorA\sthanks{Corresponding author},\, \paperauthorB,\,  \paperauthorC,\, \paperauthorD}
{\href{https://www.gaps.ssr.upm.es/}{Signal Processing Applications Group} \\ Information Processing and Telecommunications Center \\ Universidad Politécnica de Madrid, Spain\\
{\tt \href{mailto:fernando.marcos.macias@upm.es}{fernando.marcos.macias@upm.es}}
}

\begin{document}
\ifpdf 
  \DeclareGraphicsExtensions{.png,.jpg,.pdf}
\else  
  \DeclareGraphicsExtensions{.eps}
\fi


\makeatletter
\def\bstctlcite{\@ifnextchar[{\@bstctlcite}{\@bstctlcite[@auxout]}}
\def\@bstctlcite[#1]#2{\@bsphack
  \@for\@citeb:=#2\do{%
    \edef\@citeb{\expandafter\@firstofone\@citeb}%
    \if@filesw\immediate\write\csname #1\endcsname{\string\citation{\@citeb}}\fi}%
  \@esphack}
\makeatother
\bstctlcite{IEEEexample:BSTcontrol}

\maketitle

\begin{abstract}

Adaptive room equalization remains challenging under time-varying acoustic conditions and complex excitation signals, such as music. In these scenarios, classical filtered-x least mean squares (Fx-LMS) methods falter due to their rigid formulation. We present a modular differentiable digital signal processing (DDSP) framework for closed-loop adaptive room equalization that recovers Fx-LMS as a special case through automatic differentiation. The framework supports interchangeable EQ structures, response estimation methods, loss functions, and optimizers. 
Experiments with time-varying measured room impulse responses show that frequency-domain objectives provide more stable adaptation than time-domain objectives in the considered scenarios. Relative to the non-equalized response, system distance is reduced by 70\%, mel-spectral distance by 13\% (worst-case scenario) relative to the non-equalized response. We further examine how online room response estimation accuracy and frame length affect the trade-off between responsiveness and convergence stability. Overall, the framework provides a unified open-source basis for exploring synergies between classical adaptive filtering and DDSP-based optimization. 


\end{abstract}

\section{Introduction}
\label{sec:intro}

Room Equalization (RE) compensates linear distortions from playback equipment and room acoustics at the intersection of digital signal processing (DSP) and active acoustics \cite{Cecchi2017}. In practice, the transfer function of sound systems varies continuously with source-listener position, room geometry, crowd density, device deviations, and environmental factors like temperature or wind in outdoor venues \cite{mourjopoulos1985variation,hatziantoniou2004errors,Carini2012}.
These effects pose acute challenges in live-event mixing, where precomputed equalizers (EQs) fail to remain optimal as time progresses. Adaptive Room Equalization (ARE) addresses this by recasting RE as a closed-loop control problem, where EQ parameters adapt to maintain the target response amid dynamic acoustic variations.

Rather than a single canonical algorithm, ARE has evolved as a family of distinct formulations. The classical approach models the equalizer as a finite impulse response (FIR) adaptive filter, solved via filtered-x least mean squares (Fx-LMS)—essentially a specific case of the adaptive linear combiner \cite{WidrowB1981,Bjarnason1995}.
This Fx-LMS foundation has spawned numerous improvements: step-size adaptation \cite{Jariwala2017} and second-order optimization \cite{Li2016} for faster convergence, weight biasing for stability \cite{Fuster2012}, perceptual equalization to match human hearing \cite{Shi2023}, and block-based or subband schemes for efficiency and robustness \cite{Fuster2016-1,Fuster2016-2,Cecchi2011,Cecchi2021,ancona2025non}. 
In parallel, frequency-domain inversion and multipoint formulations \cite{Carini2012,Cecchi2014,Jing2021} tackle the same equalization goal from fundamentally different optimization perspectives.
Such diversity across EQ structures, loss functions, and update rules calls for a unified framework that simplifies prototyping adaptive room equalizers.

Recent advances in DDSP and deep learning have dramatically expanded EQ design possibilities beyond Fx-LMS variants. Differentiable parametric EQs now integrate with diverse neural architectures---e.g. feedforward and convolutional neural networks, and Kolmogorov--Arnold networks---for response matching and style transfer tasks \cite{Nercessian2020,Steinmetz2022,Pepe2022,Mockenhaupt2024,malek2025biquad,Parakrant2025}. Open-source differentiable audio toolkits have made these approaches practical \cite{DalSanto2025}. DDSP-based approaches have shown that equalization can be formulated as a machine learning problem, enabling flexible choices of parameterization and loss function. However, these methods have been studied mainly in offline and static settings, often explicitly excluded from adaptive applications \cite{Pepe2022}. In contrast, the present work focuses on closed-loop adaptive equalization under time-varying acoustics, using a modular differentiable formulation that also allows interchangeable loss functions and update rules but within an online control setting, bridging the conceptual gap between DDSP and adaptive filtering methodologies.



This paper introduces a DDSP framework for ARE that unifies classical adaptive filtering with differentiable signal processing. It is characterized by a fully modular pipeline with interchangeable equalizer architectures, response estimation, loss functions, and optimizers—within which classical Fx-LMS arises naturally as a special case. Frequency-domain loss functions and advanced optimizers (including homotopy-based iHAM) further improve adaptation under nonstationary musical excitation, as demonstrated experimentally. Neural extensions remain future work.

The main contributions of this work are:
\begin{itemize}
    \item a DDSP-based ARE framework using automatic differentiation in true closed-loop, subsuming classical methods including Fx-LMS;
    \item an open-source PyTorch implementation for reproducibility and framework exploration \footnote{https://github.com/fermarcosmac/DDSP-adaptive-EQ-26.git};
    \item a systematic experimental evaluation under nonstationary music and time-varying acoustics from real-world RIR measurements \cite{wang2023soundcam}.
\end{itemize}

This paper is not intended as a final practical solution for deployed live-room equalization, nor as a fully factorized comparison of all adaptive equalization design choices. Rather, its goal is to introduce a framework that makes these choices explicit and interoperable, establishes the connection to Fx-LMS, and demonstrates, in controlled time-varying simulations, that structured parametric equalization with frequency-domain objectives is a promising ARE direction under musical excitation.


\section{Proposed Framework}
\label{sec:prop-framework}

\begin{figure}[t]
\centerline{\includegraphics[scale=0.75]{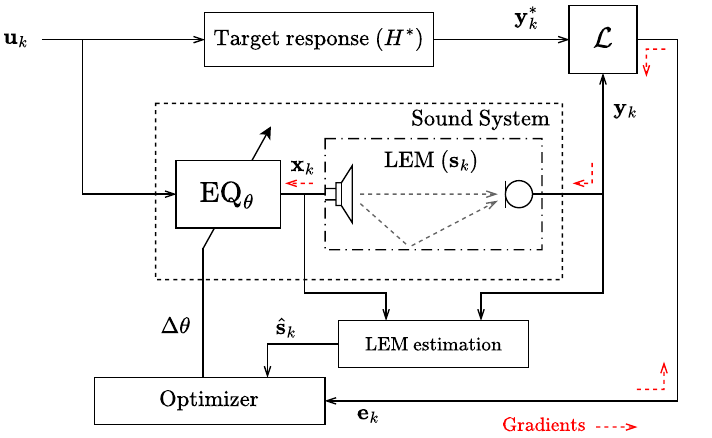}}
\caption{\label{fig:Adaptive_EQ_schematic_cropped}{Block diagram of the proposed adaptive room equalization system.}}
\end{figure}

Our ARE framework is a closed-loop controller that adapts the parameters of an EQ by minimizing the loss $\mathcal{L}$ between the equalized system response and a target response $H^*$. Figure~\ref{fig:Adaptive_EQ_schematic_cropped} shows the corresponding block diagram, in which the EQ parameters are continuously updated to compensate for time-varying linear distortions in the loudspeaker-enclosure-microphone (LEM) tuple. The framework exposes four configurable components: (i) the EQ structure parameterized by $\hat{\boldsymbol{\theta}}$, (ii) the dynamic response estimation method, (iii) the loss function $\mathcal{L}$ used to measure deviation from the target response, and (iv) the optimizer used to update the parameters. Since these design choices affect both convergence and steady-state performance, we evaluate them experimentally in this work.

Assuming a discrete-time setting that includes DAC/ADC effects, the sound system can be modeled as a slowly varying linear filter with unknown response $\mathbf{s}_k$.
The input signal $u$ is segmented into fixed-length frames $\mathbf{u}_k = \left[u(kN), \dots, u(kN+N-1)\right]^T$, which pass through the parametric EQ and the sound system to produce the measured output $\mathbf{y}_k$ at the capture device. Index $k$ numbers the time frames. This output is then compared with the target output $\mathbf{y}_k^*$, which is produced via desired response $H^*$. The latter may range from a pure delay~\cite{Fuster2012,Fuster2016-1,Fuster2016-2} to more realistic distance-dependent spectral profiles with low-frequency roll-off~\cite{Vlimki2016}. The resulting deviation is quantified by a differentiable loss $\mathcal{L}(\mathbf{y}_k,\mathbf{y}_k^*)$, ensuring end-to-end differentiability from the EQ parameters $\hat{\boldsymbol{\theta}}$ to the objective function. Parameter updates follow

\begin{equation}
\hat{\boldsymbol{\theta}}_{k+1} = \hat{\boldsymbol{\theta}}_k + \Delta\hat{\boldsymbol{\theta}}_k
\label{eq:update}
\end{equation}
\noindent leveraging the aforementioned differentiability (e.g., via gradients). Finally, for the purposes of this work, real-time sound system operation affords only one forward pass per frame $\mathbf{u}_k$, so no frame overlap is permitted and exactly one parameter update is performed per frame. This imposes a fundamental tradeoff in frame length selection between update rate and spectral resolution, which is analyzed empirically in the ablation study (Section \ref{sec:ablations}).


Hereafter, we detail the specific implementations of these modular components, towards the experimental validation of this work.



\subsection{Differentiable Parametric Equalizer}
We focus on the standard parametric room EQ formulation based on cascaded biquad filters \eqref{eq:biquad_system_fcn}, which are computationally efficient and readily available in audio processing devices and software. In this section only, we will drop the time frame index $k$ for the sake of clarity. The $m$-th biquad filter is parametrized equivalently by its filter coefficients $\left(a_{i,m},b_{i,m}\right)$, its poles and zeros, or its EQ parameters $\boldsymbol{\theta}_m$: type (e.g., high/low shelf, peaking), center/cutoff frequency ($f_m$), gain ($g_m$) and quality factor ($Q_m$). That is, $\boldsymbol{\theta}_m=[f_m, g_m, Q_m] \in \mathbb{R}^3$. We adopt the latter parametrization due to its widespread use and interpretability. The transfer function of the $m$-th biquad filter is:
\begin{equation}
\label{eq:biquad_system_fcn}
    H_k(z;\boldsymbol{\theta}_m) = \frac{b_{0,m}+b_{1,m}z^{-1}+b_{2,m}z^{-2}}{a_{0,m}+a_{1,m}z^{-1}+a_{2,m}z^{-2}}
\end{equation}
The (differentiable) mapping between biquad coefficients and EQ parameters $\boldsymbol{\theta}_m = [f_m,\,g_m,\,Q_m]$ is detailed in \cite{Nercessian2020}. Accordingly, all filter coefficients depend on $\boldsymbol{\theta}_m$, although this dependence is omitted in the notation \eqref{eq:biquad_system_fcn} for clarity. While time-domain implementation of the IIR filter is both feasible and differentiable, its recursive structure entails the well-known limitations of backpropagation through time \cite{Steinmetz2022}. Thus, a FIR approximation based on the frequency sampling method \cite{DalSanto2025} is preferred for differentiation, while the actual parametric EQ may remain IIR and efficient.

The total frequency response of the $M$-band parametric EQ is 
\begin{equation*}
\label{eq:paramEQ_system_fcn}
    H_{EQ}(e^{j\omega};\hat{\boldsymbol{\theta}}) = G\cdot\textstyle\prod_{m=1}^{M} H_m(e^{j\omega};\boldsymbol{\theta}_m)
\end{equation*}
where $G$ is an overall gain and $\hat{\boldsymbol{\theta}}$ is the complete parameter vector obtained by concatenating the $M$ biquad parameter vectors $\boldsymbol{\theta}_m$ and $G$. 
The output of the EQ block is given by %
$X(e^{j\omega};\hat{\boldsymbol{\theta}}) = U(e^{j\omega}) \cdot H_{EQ}(e^{j\omega};\hat{\boldsymbol{\theta}})$
%
where the product is evaluated at a discrete set of frequencies. In any ARE application, the signal at the EQ output $\mathbf{x}(\hat{\boldsymbol{\theta}})$ traverses the sound system with unknown, slowly varying response $\textbf{s}$, producing output $\mathbf{y}(\hat{\boldsymbol{\theta}}) = \mathbf{x}*\mathbf{s}\,(\hat{\boldsymbol{\theta}})$. This linear model provides a differentiable parameters-to-output mapping $\hat{\boldsymbol{\theta}} \mapsto \mathbf{y}(\hat{\boldsymbol{\theta}})$. In practice, the output is not computed; it is measured at the listening position. The 
mapping is just needed to compute the derivative information.

\subsection{Loss Function}
The deviation between the captured output frame $\mathbf{y}_k$ and the target output $\mathbf{y}_k^*$ is quantified by a loss function $\mathcal{L}(\mathbf{y}_k, \mathbf{y}_k^*;\hat{\boldsymbol{\theta}})$. The choice of $\mathcal{L}$ is critical, as it defines the optimization criterion for the EQ parameters $\hat{\boldsymbol{\theta}}$. A baseline time-domain loss (TD-MSE) minimizes the mean squared error between the output $\mathbf{y}_k$ and target $\mathbf{y}_k^*$ waveforms.
%
%
However, frame-to-frame variability in the time-domain---particularly for nonstationary signals such as speech or music---poses significant convergence challenges, as observed in Fx-LMS algorithms \cite{WidrowB1981}. For longer frames, greater flexibility emerges in the design of $\mathcal{L}$. In particular, frequency-domain loss functions \cite{shynk2002frequency,malek2025biquad,Nercessian2020,Jing2021} and subband variants of Fx-LMS \cite{Cecchi2011,Cecchi2021,ancona2025non} yield improved robustness to nonstationarity. Moreover, online estimates of the equalized system's magnitude response---obtained, for example, via frequency-domain deconvolution \cite{Kirkeby1998}---enable direct comparison to a desired target response ---see \textit{response loss} in \cite{malek2025biquad,Nercessian2020}. Therefore, in our experimental setup (Section \ref{sec:exp-setup}) we use:
\begin{equation}
\label{eq:FD-MSE}
    \text{FD-MSE}(\mathbf{y}_k,H^*) =  \frac{1}{N}\sum_{n=1}^{N} \left( \left| \frac{Y_k}{U_k}(e^{j\omega_n})\right| - \left| H^*(e^{j\omega_n}) \right| \right)^2
\end{equation}
\noindent where $Y_k(e^{j\omega})$ and $U_k(e^{j\omega})$ are the $N$-point DFTs of the output and input frames, respectively. The framework supports combining multiple loss functions \cite{Nercessian2020} and incorporating perceptually motivated control criteria, such as warped frequency sampling \cite{Carini2012,Shi2023}. Our open-source implementation allows experimenting with alternative loss functions in various scenarios and the same applies to the optimizer block (see next section). 

\subsection{Optimizer}
\label{subsec:optimizer}

We address the ARE problem using derivative-based iterative optimization algorithms, which may only perform one forward pass per input frame $\mathbf{u}_k$. 
To illustrate the flexibility of the proposed framework, we evaluate several derivative-based update rules, including SGD \cite{nocedal2006numerical}, Adam \cite{Kingma2014}, Newton \cite{nocedal2006numerical}, and a homotopy-analysis-based method (iHAM) \cite{Liao1992,Zhao2013,Liu2023}. As mentioned before, the purpose of this comparison is not to claim a universally superior optimizer, but rather to show that the framework can accommodate both standard and non-standard update mechanisms within the same differentiable control loop. In this sense, iHAM is included as an exploratory example of a higher-order alternative, suggested to avoid local minima more effectively than gradient-based approaches \cite{nocedal2006numerical}, which is attractive given the non-convexity of EQ matching \cite{Nercessian2020}.
\vspace{6pt}

\noindent\textbf{SGD} performs a single step in the negative gradient direction:
\begin{equation}
    \hat{\boldsymbol{\theta}}_{k+1} = \hat{\boldsymbol{\theta}}_k - \eta_k \nabla_{\hat{\boldsymbol{\theta}}}\mathcal{L}(\mathbf{y}_k, \mathbf{y}_k^*;\hat{\boldsymbol{\theta}}_k)
\end{equation}

\noindent\textbf{Newton} uses second-order information via the Hessian inverse for quadratic convergence near stationary points:
\begin{equation}
    \hat{\boldsymbol{\theta}}_{k+1} = \hat{\boldsymbol{\theta}}_k - \left[ \nabla_{\hat{\boldsymbol{\theta}}}^2 \mathcal{L}(\mathbf{y}_k, \mathbf{y}_k^*;\hat{\boldsymbol{\theta}}_k) \right]^{-1} \nabla_{\hat{\boldsymbol{\theta}}}\mathcal{L}(\mathbf{y}_k, \mathbf{y}_k^*;\hat{\boldsymbol{\theta}}_k) 
\end{equation}

\noindent\textbf{Adam} maintains exponential moving averages of the gradient ($\mathbf{m}_k$) and its squared magnitude ($\mathbf{v}_k$), yielding adaptive per-parameter learning rates \cite{Kingma2014}:
\begin{align}
\mathbf{m}_{k+1} &= \beta_1 \mathbf{m}_{k} + (1 - \beta_1) \nabla_{\hat{\boldsymbol{\theta}}} \mathcal{L}(\hat{\boldsymbol{\theta}}_{k}) \nonumber\\ 
\mathbf{v}_{k+1} &= \beta_2 \mathbf{v}_{k} + (1 - \beta_2) [\nabla_{\hat{\boldsymbol{\theta}}} \mathcal{L}(\hat{\boldsymbol{\theta}}_{k})]^2 \nonumber\\
\hat{\mathbf{m}}_{k+1} &= \frac{\mathbf{m}_{k+1}}{1 - \beta_1^{k+1}}, \quad \hat{\mathbf{v}}_{k+1} = \frac{\mathbf{v}_{k+1}}{1 - \beta_2^{k+1}} \nonumber\\
\hat{\boldsymbol{\theta}}_{k+1} &= \hat{\boldsymbol{\theta}}_{k} - \eta \frac{\hat{\mathbf{m}}_{k+1}}{\sqrt{\hat{\mathbf{v}}_{k+1}} + \epsilon}
\end{align}

\noindent\textbf{iHAM} is less common in the audio signal processing literature and is therefore described in slightly more detail. Rather than stepping in the gradient direction, it minimizes the loss by seeking a root of
\begin{equation}
\label{eq:loss_root_eq}
    \mathcal{L}(\mathbf{y}_k, \mathbf{y}_k^*;\hat{\boldsymbol{\theta}}_k)-\varepsilon_0=0 \nonumber
\end{equation}
where $\varepsilon_0$ is a hyperparameter representing the irreducible error. Starting from $\hat{\boldsymbol{\theta}}_k$, iHAM constructs a \emph{linearized approximation} of the loss surface around $\hat{\boldsymbol{\theta}}_k$ using the Jacobian $D_{\hat{\boldsymbol{\theta}}_k}\mathcal{L}$, then continuously deforms this approximation toward the true loss via the \emph{zeroth-order deformation equation}:
\begin{equation}
\label{eq:ZODE}
    (1-q)\,D_{\hat{\boldsymbol{\theta}}_k}\mathcal{L} \, (\phi(q)-\hat{\boldsymbol{\theta}}_k) = -q\,\eta \, \left( \mathcal{L}(\phi(q))-\varepsilon_0 \right) \nonumber
\end{equation}
\noindent where $\eta$ is a convergence control hyperparameter, $q \in [0,1]$ is the embedding parameter, and $\phi(q)$ interpolates from $\hat{\boldsymbol{\theta}}_k$ (at $q=0$) to the solution $\hat{\boldsymbol{\theta}}^*$ (at $q=1$). Approximating $\phi(q)$ by a truncated Maclaurin series and setting $q=1$ yields:
\begin{equation}
    \hat{\boldsymbol{\theta}}_{k+1} = \hat{\boldsymbol{\theta}}_k + \textstyle\sum_{j=1}^{J} \hat{\boldsymbol{\theta}}_{k,j} \, = \hat{\boldsymbol{\theta}}_k + \Delta \hat{\boldsymbol{\theta}}
\end{equation}
where corrections $\hat{\boldsymbol{\theta}}_{k,j}$ come from solving linear \textit{higher-order deformation equations} recursively. The truncation order $J$ controls the accuracy--complexity trade-off, giving rise to the iHAM-$J$ family. At $J=1$, the update reduces to the linear system
\begin{equation}
\label{eq:iHAM-1_update}
    \nabla_{\hat{\boldsymbol{\theta}}_{k}}\mathcal{L} \cdot \Delta\hat{\boldsymbol{\theta}} = -\eta \, \left( \mathcal{L}(\hat{\boldsymbol{\theta}}_{k}) - \varepsilon_0 \right) \nonumber
\end{equation}
\noindent solved for $\Delta\hat{\boldsymbol{\theta}}$, followed by $\hat{\boldsymbol{\theta}}_{k+1} = \hat{\boldsymbol{\theta}}_k + \Delta\hat{\boldsymbol{\theta}}$. iHAM naturally encompasses other optimizers such as Newton's method \cite{Liao1992} and SGD, given the considerable flexibility of its original formulation. Its full generality is beyond the scope of this paper; the particular instance considered here is meant to illustrate the prototyping capabilities of the proposed ARE framework.


\begin{figure}[t]
\centerline{\includegraphics[scale=0.5]{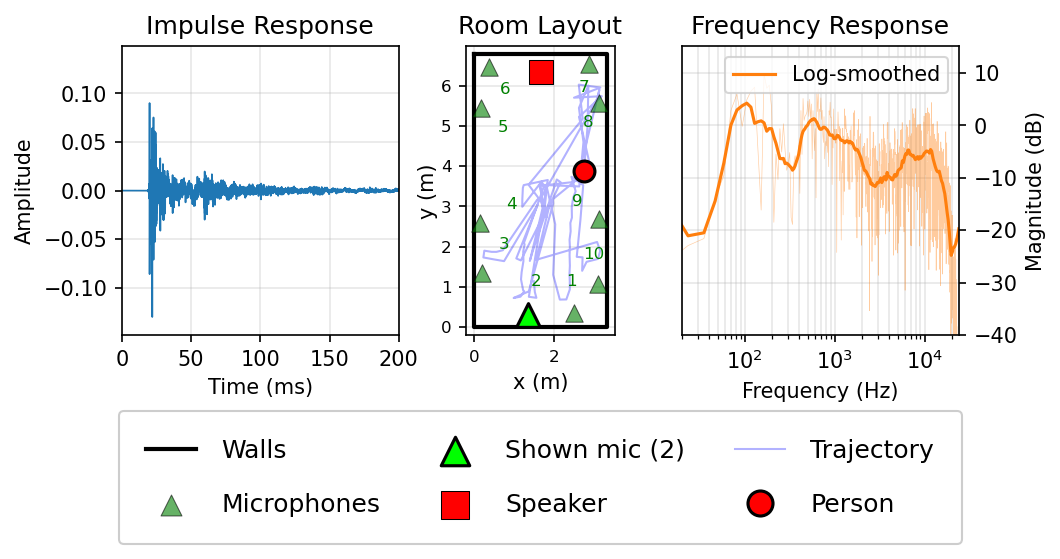}}
\caption{\label{fig:human1_animation_frame_mic2_dp0}{SoundCam conference room layout and sample RIR measurement \cite{wang2023soundcam}.}}
\end{figure}

\begin{figure*}[t!]
\centering
\includegraphics[scale=0.63]{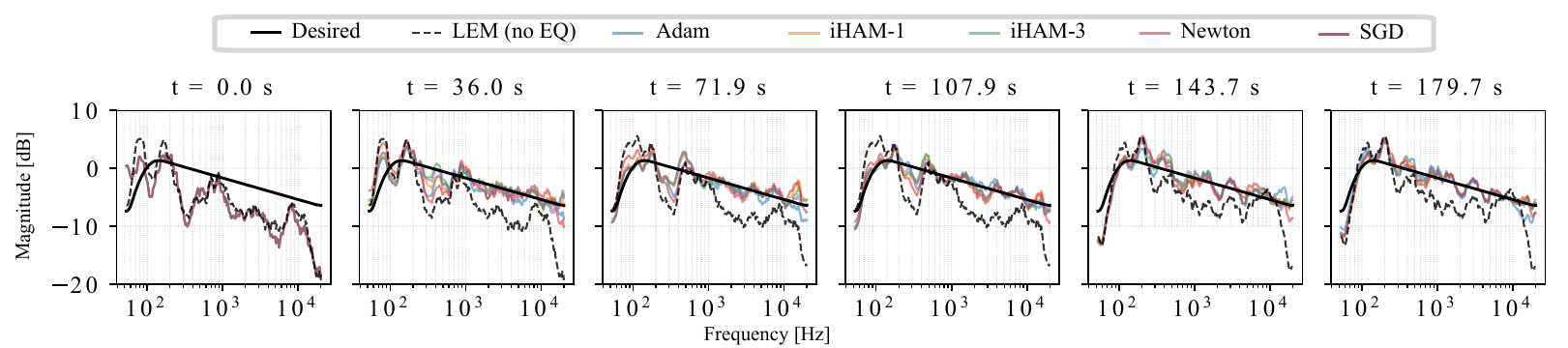}
\caption{Example of ARE experiment for different optimizers, FD-MSE loss and a musical excitation signal. As time progresses, the magnitude response of the sound system (dashed line) changes, and the parametric EQ is continuously adapted to create a combined response more similar to target (solid line). Optimizers: Adam (blue), iHAM-1 (orange), iHAM-3 (green), Newton (red), and SGD (purple).}
\label{fig:EQ_response_example}
\end{figure*}

\subsection{Derivative Flow}
\label{subsec:derivative_flow}
We employ automatic differentiation (AD) throughout the full signal chain. Whether using forward-mode AD or reverse-mode AD (i.e., \textit{backpropagation}), the derivative of each operation in the computational graph---from the parameters $\hat{\boldsymbol{\theta}}$ to the loss $\mathcal{L}$---must be evaluated. In particular, gradients must propagate through both the parametric EQ and the model of the sound system. All operations within the parametric EQ and the loss function are explicitly designed to be differentiable. Since the sound system is modeled as a linear and slowly time-varying filter, the Jacobian of the output frame with respect to the input frame is given by the convolution operator induced by the impulse response $\mathbf{s}_k$. Consequently, gradient backpropagation through the sound system obeys

%
\begin{equation*}
    \nabla_{\mathbf{x}_k}\mathcal{L} =  \nabla_{\mathbf{y}_k} \mathcal{L} \ast \mathbf{s}'_k
\end{equation*}

\noindent where $\ast$ stands for the convolution operator and $\mathbf{s}'_k$ represents the time-reversed sound system response $\mathbf{s}_k$. This is a standard result from linear systems theory \cite{Cartan1971}. Since $\mathbf{s}_k$ is unknown \textit{a priori}, an online estimate $\hat{\mathbf{s}}_k$ approximates this gradient flow.  A fast and robust way to gain this estimate from $\mathbf{x}_k$ and $\mathbf{y}_k$ is frequency-domain regularized deconvolution \cite{Kirkeby1998}.
This approach mirrors the ``filtered-x'' signal computation in Fx-LMS algorithms and enables end-to-end gradient flow for the AD computation.


\subsection{Fx-LMS as a Special Case}
\sloppy
Fx-LMS is recovered under three assumptions: (i) the EQ is FIR, so
$\hat{\boldsymbol{\theta}}_k$ is its parameter vector, numerically equal to its impulse
response; (ii) adaptation uses single-sample frames; and (iii) the loss is the
instantaneous squared error,
$\mathcal{L}(\mathbf{y}_k;\hat{\boldsymbol{\theta}}_k)
= \frac{1}{2}(\mathbf{y}_k-\mathbf{y}_k^*)^2$ \cite{WidrowB1981,Bjarnason1995}.
Under these assumptions, the SGD gradient is
\begin{align}
    \nabla_{\hat{\boldsymbol{\theta}}_k}\mathcal{L}
    &= \nabla_{\hat{\boldsymbol{\theta}}_k}
    \frac{1}{2}\left(\mathbf{u}_k*\hat{\boldsymbol{\theta}}_k*
    \mathbf{s}_k-\mathbf{y}_k^*\right)^2 \nonumber\\
    &= \mathbf{e}_k(\mathbf{s}'_k*\mathbf{u}_k)
    \approx \mathbf{e}_k(\hat{\mathbf{s}}'_k*\mathbf{u}_k),
    \nonumber
\end{align}
which yields the Fx-LMS update
\begin{equation}
    \hat{\boldsymbol{\theta}}_{k+1}
    = \hat{\boldsymbol{\theta}}_k
    - \eta(\mathbf{y}_k-\mathbf{y}_k^*)
    (\hat{\mathbf{s}}'_k*\mathbf{u}_k).
\end{equation}
Thus, the filtered-x signal is simply the gradient obtained by backpropagating
through the estimated sound system, as derived in
Section~\ref{subsec:derivative_flow}. The proposed framework relaxes the three
assumptions by allowing non-FIR or parametric EQs, framewise or frequency-domain
losses, and general differentiable optimizers, while preserving this same
source--EQ--sound-system--error computational graph. Fx-LMS is therefore the
FIR, single-sample, squared-error limit of the proposed differentiable
formulation, not a separate update rule.

\section{Experimental Setup}
\label{sec:exp-setup}
This section describes the datasets, signal chain, algorithmic configurations, and evaluation methodology used to assess the proposed adaptive room equalization (ARE) framework. The experiments evaluate both the convergence behaviour and the final equalization accuracy across different optimizer and loss-function configurations under time-varying acoustic conditions. We also test the performance of two classical adaptive filtering algorithms---Fx-LMS and filtered-x frequency domain adaptive filtering (Fx-FDAF) \cite{shynk2002frequency}---in the same scenario. The code, setup and the exact lists of RIRs and music tracks used in the experiments are available in the accompanying repository to support full reproducibility. The implementation is intended to facilitate further analysis of the method’s sensitivity to parameter settings.

\subsection{Datasets \& simulated scenarios}
Room impulse responses are drawn from the SoundCam dataset \cite{wang2023soundcam} 
(Conference Room subset), providing 48~kHz measurements at ten microphone positions 
for both empty and occupied room configurations ---see Figure~\ref{fig:human1_animation_frame_mic2_dp0}. 
These recordings capture realistic spatial variations in room transfer functions due to listener 
position and single-occupant placement.


We evaluate two scenarios: (i) changing the listener position by transitioning between RIRs measured at different microphone locations, and (ii) keeping the listener fixed while varying the single occupant’s position. Temporal evolution is controlled via linear interpolation of the frequency responses to generate transitions from abrupt (1 s) to slow drift (30 s). The varying-listener case produces the largest structural changes across frequency and is therefore treated as the worst-case scenario for controller evaluation.

For exciting the system, we used ten full-track mixes from MedleyDB \cite{bittner2014medleydb}, cropped to 180~s and resampled from 44.1~kHz to 48~kHz. The selection spans a range of genres (indie ×4, jazz ×1, pop ×1, ballad ×1, metal ×1, ambient ×1, classical ×1) to expose the controller to diverse spectral and temporal statistics representative of practical playback material.


\begin{figure*}[!t]
\centering
\includegraphics[scale=0.6]{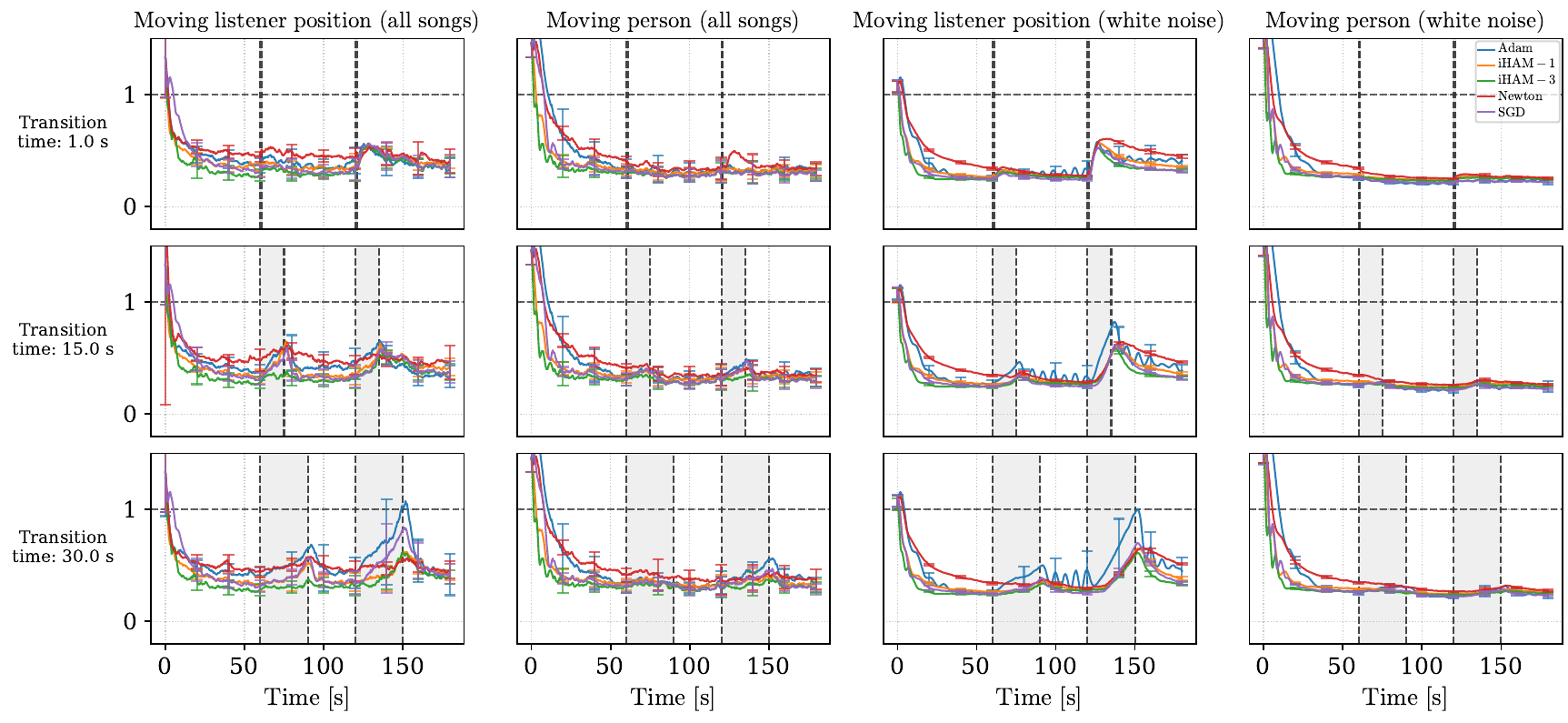}
\caption{Validation error ($D_{\text{rel}}$) across 3 minutes of playback material, for different changes in the LEM responses and transition speeds between them. Average and standard deviation values for $D_{\text{rel}}$ are depicted both for all 10 songs and 10 independent white noise runs. Shaded areas mark transitions between SoundCam room responses; in white regions, responses are fixed. The different optimizers in the control loop---Adam (blue), iHAM-1 (orange), iHAM-3 (green), Newton (red) and SGD (purple)---were set to optimize the FD-MSE loss.}
\label{fig:all_songs_moving_position}
\end{figure*}

\begin{table*}[h]
\centering
\footnotesize
\setlength{\tabcolsep}{1.5pt}
\renewcommand{\arraystretch}{0.95}
\caption{Average metrics for transition times (1\,s, 15\,s, 30\,s) across all songs, in moving-listener scenario. Fx-LMS, no convergence (NC).}
\label{tab:transition_metrics}

\resizebox{\textwidth}{!}{%
\begin{tabular}{lcccccccccccccccccccccc}
\toprule
 & \multicolumn{7}{c}{\textbf{Transition 1 s}} & \multicolumn{7}{c}{\textbf{Transition 15 s}} & \multicolumn{7}{c}{\textbf{Transition 30 s}} \\
\cmidrule(lr){2-8} \cmidrule(lr){9-15} \cmidrule(lr){16-22}
\textbf{}
& \textbf{PEAQ} & \textbf{SI-SDR} & \textbf{STFT} & \textbf{MSD} & \textbf{SCE} & \textbf{RMSE} & \textbf{LUFS}
& \textbf{PEAQ} & \textbf{SI-SDR} & \textbf{STFT} & \textbf{MSD} & \textbf{SCE} & \textbf{RMSE} & \textbf{LUFS}
& \textbf{PEAQ} & \textbf{SI-SDR} & \textbf{STFT} & \textbf{MSD} & \textbf{SCE} & \textbf{RMSE} & \textbf{LUFS} \\
\midrule
\textbf{None}
& -1.90 & -25.52 & 1.25 & 4.73 & 260.12 & 0.18 & \textbf{0.68}
& -1.90 & -25.52 & 1.25 & 4.73 & 260.12 & 0.18 & \textbf{0.68}
& -1.90 & -25.52 & 1.25 & 4.73 & 260.12 & 0.18 & \textbf{0.68} \\
\textbf{Fx-LMS}
& NC & NC & NC & NC & NC & NC & NC
& NC & NC & NC & NC & NC & NC & NC
& NC & NC & NC & NC & NC & NC & NC \\
\textbf{Fx-FDAF}
& -1.90 & -24.66 & 1.71 & 4.82 & 500.91 & 0.16 & 3.03
& -1.90 & \textbf{-24.66} & 1.71 & 4.82 & 500.91 & 0.16 & 3.03
& -1.90 & \textbf{-24.66} & 1.71 & 4.82 & 500.91 & 0.16 & 3.03 \\
\textbf{SGD}
& -1.90 & -26.11 & 4.17 & 4.26 & 231.66 & \textbf{0.15} & 6.46
& -1.90 & -25.32 & 4.25 & 4.30 & 248.36 & \textbf{0.15} & 6.79
& -1.90 & -29.83 & 4.27 & 4.40 & 269.24 & \textbf{0.15} & 7.05 \\
\textbf{Adam}
& -1.90 & \textbf{-24.48} & \textbf{1.20} & 4.23 & 215.70 & 0.17 & 1.01
& -1.90 & -25.78 & \textbf{1.19} & 4.27 & 221.34 & 0.17 & 0.91
& -1.90 & -28.42 & 1.44 & 4.42 & 288.73 & 0.17 & 1.80 \\
\textbf{iHAM-1}
& -1.90 & -26.00 & 1.76 & \textbf{4.18} & 221.97 & 0.17 & 3.44
& -1.90 & -26.33 & 1.98 & 4.22 & 233.63 & 0.17 & 3.89
& -1.90 & -28.16 & 1.89 & 4.24 & 240.00 & 0.16 & 3.65 \\
\textbf{Newton}
& -1.90 & -25.47 & 1.30 & 4.45 & \textbf{206.41} & 0.17 & 1.56
& -1.90 & -26.94 & 1.58 & 4.51 & \textbf{220.51} & 0.17 & 1.68
& -1.90 & -29.23 & \textbf{1.20} & 4.50 & \textbf{213.31} & 0.17 & 0.95 \\
\textbf{iHAM-3}
& -1.90 & -30.31 & 5.41 & 4.19 & 235.04 & \textbf{0.15} & 8.60
& -1.90 & -27.43 & 5.63 & \textbf{4.12} & 237.94 & \textbf{0.15} & 9.32
& -1.90 & -30.85 & 5.55 & \textbf{4.20} & 250.87 & \textbf{0.15} & 9.18 \\
\bottomrule
\end{tabular}%
}
\end{table*}

\subsection{Implementation details}

The processing chain in Figure~\ref{fig:Adaptive_EQ_schematic_cropped} is implemented in PyTorch for end-to-end automatic differentiation through all components. The frequency range of equalization is $50-20,000$ Hz. Audio is processed in non-overlapping frames of 8192 samples ($\approx170$~ms at 48~kHz), yielding 6~Hz spectral resolution and a 6~Hz update rate. This frame size offers a favorable compromise between LEM estimation accuracy, controller responsiveness, computational efficiency, and perceptual audio continuity \cite{Cecchi2014}; this was empirically validated in the ablation study---see Section~\ref{sec:ablations}.

Per-frame LEM response estimates use regularized frequency-domain deconvolution \cite{Kirkeby1998}, exponentially smoothed (5\% new / 95\% historical) to stabilize estimation against abrupt changes at the cost of slower adaptation. 
For simulation, a custom PyTorch function applies ground-truth LEM impulse responses during the forward pass while computing gradients via online LEM estimates in the backward pass---faithfully modeling real-world scenarios.




The differentiable parametric equalizer uses the \textit{dasp-pytorch}~\cite{Steinmetz2022} 
package to implement a cascade of seven biquad filters: a low-shelf (20--2000~Hz), five 
peaking filters covering 80--500, 80--2000, 2000--8000, 8000--12000, and 12000--23000~Hz, 
and a high-shelf (4000--23000~Hz). Shelf gains span $[-20, 20]$~dB; peaking gains 
$[-20, 10]$~dB; all $Q$ factors $[0.1, 2.0]$; and the global gain $G \in [-24, 24]$~dB. 
Initial parameters are set to the midpoint of each frequency and $Q$ range, with all 
gains at 0~dB.

The target response consists of a pure delay combined with a magnitude response exhibiting low-frequency roll-off and distance-dependent spectral decay (see Figure \ref{fig:EQ_response_example}), which is precomputed following the procedure described in \cite{matlabEQexample}. This target response corresponds to an ideal anechoic environment, with the roll-off meant to avoid equipment damage.  

To ensure a fair comparison across optimization methods, we evaluate our proposed framework using different modern gradient-based optimizers and against established adaptive filtering baselines. Specifically, we compare SGD, Adam, Newton (with damping), and iHAM---all employing the same differentiable 7-biquad parametric equalizer described above---against the classical Fx-LMS and Fx-FDAF baselines. Both classical baselines use long 2048-tap FIR equalizers  \cite{Cecchi2014,shynk2002frequency}, SGD optimization (via closed-form update rules), and rely on estimates of the LEM response for gradient computation. Fx-LMS uses time-domain MSE with per-frame updates, whereas Fx-FDAF uses a block-based frequency-domain formulation closer to ours, yet less flexible. This setup isolates the impact of optimization strategy and model flexibility while maintaining identical LEM estimation, frame processing, and experimental conditions for all methods.
The optimizer hyperparameters were selected empirically for stable behavior across all evaluated scenarios according to criteria in Section \ref{subsec:evaluation}: SGD uses $\eta = 5\cdot10^{-3}$; Adam uses $\eta = 5\cdot10^{-3}$, 
$\beta_1 = 0.9$, $\beta_2 = 1-10^{-5}$, $\epsilon = 10^{-8}$; Newton uses damping 
$\lambda = 10^{3}$; and iHAM uses, $\eta = 5\cdot10^{-3}$, 
$\varepsilon_0 = 1$. A comprehensive, formally matched hyperparameter search remains an important avenue for future work. All experiments were run on an AMD Ryzen 7 9800X3D 8-Core Processor (4.70 GHz) and an NVIDIA GeForce RTX 5090 GPU.

\subsection{Evaluation}
\label{subsec:evaluation}
The \emph{normalized relative system distance}
\begin{equation*}
    D_{\text{rel}} = \frac{\| |H_{\text{eq}}| - |H^*| \|_1}{\| |H_{\text{room}}| - |H^*| \|_1}
\end{equation*}
extends the classic system distance \cite{goetze2008decoupled,Fuster2012}---also used as a loss in deep learning EQ methods \cite{malek2025biquad}---by normalizing against unprocessed room distortion, $|H_{\text{room}}|$. It measures \emph{relative spectral-energy improvement} ($D_{\text{rel}}=1$ indicates no correction; $D_{\text{rel}}=0$ perfect equalization), fosters cross-environment comparability. 
The L1-magnitude formulation prioritizes average spectral balance over frequency outliers, aligning with perceptual equalization goals. We report temporal $D_{\text{rel}}$ trajectories and summary statistics across excitation signals to assess convergence and robustness. 

Following prior work \cite{Steinmetz2022,Parakrant2025}, we complement $D_{\text{rel}}$ with established metrics spanning perceptual quality (PEAQ: model-based MOS prediction; SI-SDR: scale-invariant signal-to-distortion ratio), spectral alignment (multi-resolution STFT error; mel-spectral distance with 1024-sample window; spectral centroid: first-moment preservation), time-domain fidelity (RMSE), and loudness consistency (LUFS difference), evaluated over 150~s (after 30~s of warmup) through time-varying acoustic conditions.

\section{Results}
\label{sec:results}

Figure~\ref{fig:all_songs_moving_position} shows $D_{\text{rel}}$ trajectories for all optimizers using the FD-MSE loss \eqref{eq:FD-MSE} across music (columns 1-2) and white noise (columns 3-4) excitations. TD-MSE optimizations failed to converge ($D_{\text{rel}}>1.0$) across all tested configurations, confirming prior observations that time-domain loss functions struggle with musical nonstationarity \cite{WidrowB1981,shynk2002frequency}.

All FD-MSE configurations achieve $D_{\text{rel}}<1.0$, indicating spectral improvement over unprocessed room responses. However, substantial variance across the 10 music tracks reveals sensitivity to excitation statistics. Standard deviations span 0.1--0.2 $D_{\text{rel}}$ during steady-state periods. iHAM-3 exhibits the lowest average $D_{\text{rel}}$ across 1~s, 15~s, and 30~s transitions in the moving-listener (worst-case) scenario, followed by iHAM-1 and Newton. Adam shows instability in 3 out of 10 music tracks during 30~s transitions, while SGD converges reliably but reaches higher steady-state error floors---$\Delta D_{\text{rel}}\approx0.05$ higher than \ iHAM-3. The observed instability around smooth transitions is explored in Section \ref{sec:ablations}.

Table~\ref{tab:transition_metrics} reports metrics for 150~s scenarios (after 30~s warmup) across moving-listener transitions. iHAM-3 achieves lowest mel-spectral distance (MSD: 4.12, 15~s transition) among tested optimizers, consistent with FD-MSE optimization. Newton minimizes spectral centroid error (SCE: 213--220~Hz), suggesting better preservation of spectral balance despite higher $D_{\text{rel}}$. Time-domain metrics (SI-SDR $\approx$-25~dB, RMSE $\approx$0.15) show minimal change post-equalization due to phase-preserving target design. Informal listening tests confirmed audible improvements over unprocessed responses; MUSHRA evaluation remains future work.

\begin{figure}[t]
\centerline{\includegraphics[scale=0.65]{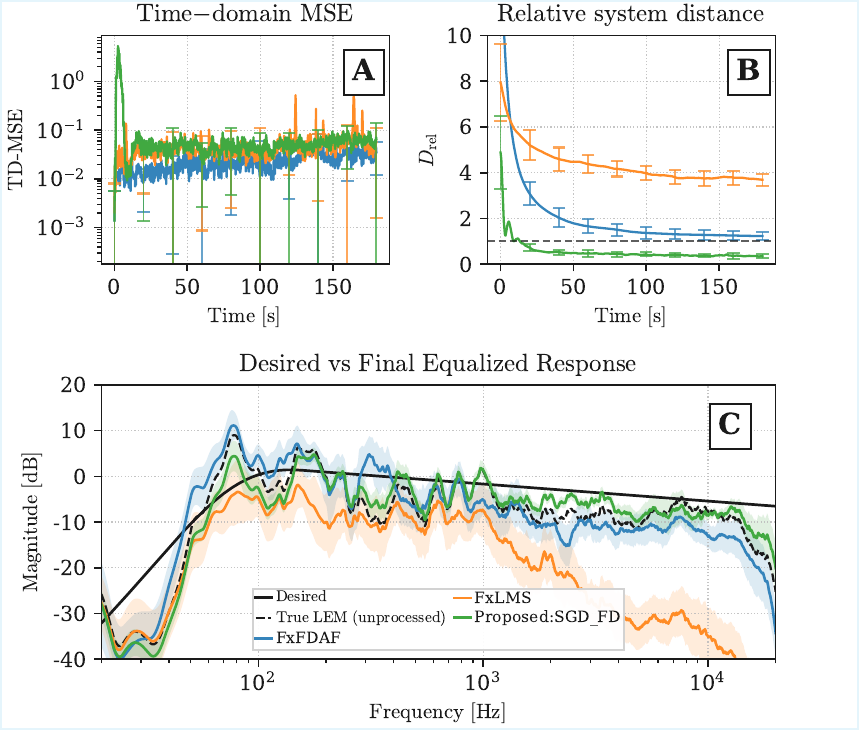}}
\caption{\label{fig:comparison_FIR_classical}{ARE performance comparison between a configuration (SGD, FD-MSE, parametric EQ) of the proposed framework with the classical Fx-LMS (time-domain) and Fx-FDAF (frequency-domain) methods, across all music tracks. Summary statistics (average, std) are provided for time-domain error (A), system distance $D_{\text{rel}}$ (B) and final equalized magnitude response (C).}}
\end{figure}

Classical baseline Fx-LMS failed to converge for the modified pyaec implementation\footnote{https://github.com/ewan-xu/pyaec v1.0.1 (PyPI).} in the time-varying moving-listener scenario (Table~\ref{tab:transition_metrics}, marked "NC"). From the same library, Fx-FDAF (2048-tap FIR equalizers) converged but yielded higher $D_{\text{rel}}$ (Fig.~\ref{fig:all_songs_moving_position} and Fig.~\ref{fig:comparison_FIR_classical}) despite its frequency-domain formulation. The proposed parametric EQ (22 parameters) outperformed these FIR baselines (2048 taps), highlighting advantages of structured equalization under parameter count constraints.

Per-frame computation times for stationary excitations (convergent for all methods and running on RTX 5090, see Table~\ref{tab:compute_time}) remain below 170~ms frame duration. Only Fx-LMS exceeds frame time (201~ms). First-order methods average 20~ms (82\% headroom), while higher-order methods (Newton, iHAM-3) require 140--141~ms (17\% headroom). Compute times show tight ranges (min-max variation <2~ms), indicating negligible OS jitter impact under these conditions. These measurements exclude I/O latency and OS jitter, limiting a full real-time assessment.

The proposed frame-based framework with FD-MSE demonstrates robust spectral tracking across acoustic transitions and music excitations, with iHAM-3 offering the best performance ($D_{\text{rel}}$ control) among evaluated configurations. Baseline comparison underscores sensitivity of FIR methods to long impulse responses and nonstationary inputs characteristic of the tested scenarios.

\begin{figure*}[t]
\centering
\includegraphics[scale=0.62]{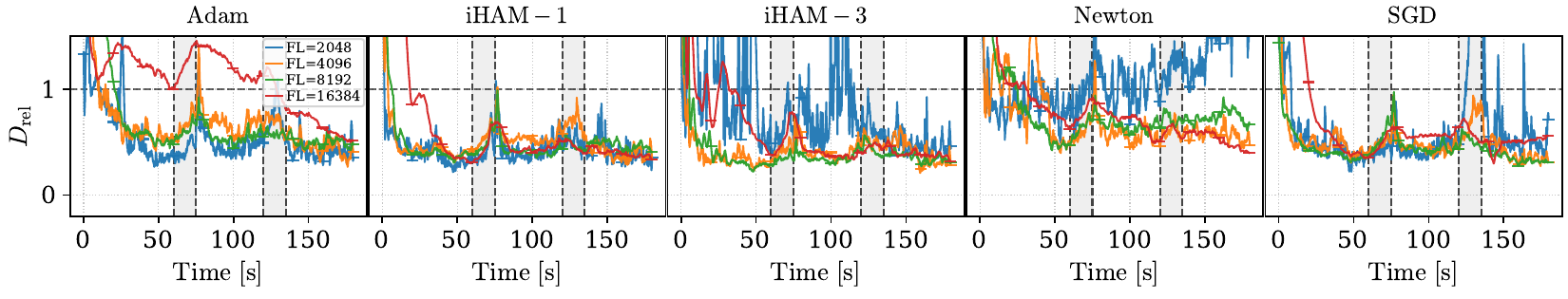}
\caption{Comparison of frame sizes (different colors) in control experiments for all optimizers and one musical excitation signal (no averaging so adaptation dynamics are portrayed in full detail). Frame sizes range from 2,048 to 16,384 samples. The transition time (shaded area) is 15 s and all optimizers use the FD-MSE loss.}
\label{fig:frame_size_analysis}
\end{figure*}

\begin{figure}[t]
\centerline{\includegraphics[scale=0.6]{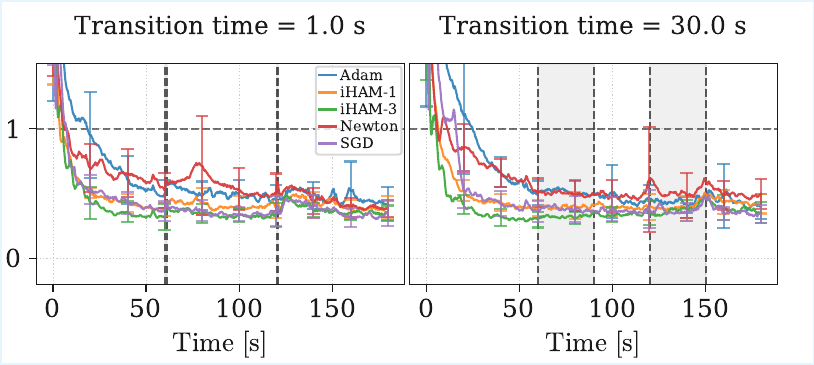}}
\caption{\label{fig:true_LEM_ablation}{Validation error ($D_{\text{rel}}$) across 3 minutes of musical input and changes in listener position. In this case, the ground-truth LEM is used for gradient computation instead of the online estimate needed for real-world ARE applications.}}
\end{figure}

\begin{table}[h]
\centering
\caption{Per-frame compute time for optimizers and FIR baselines}
\label{tab:compute_time}
\begin{tabular}{l c c c}
\hline
\textbf{Optimizer} & \textbf{Mean (ms)} & \textbf{Min (ms)} & \textbf{Max (ms)} \\
\hline
Fx-LMS & 200.83 & 199.53 & 201.91 \\
Fx-FDAF & 16.57 & 16.23 & 16.66 \\
SGD    & 19.86  & 19.79 & 19.97 \\
Adam   & 19.96  & 19.88 & 20.02 \\
iHAM-1 & 20.49  & 20.22 & 22.15 \\
Newton & 140.15 & 139.90 & 140.96 \\
iHAM-3 & 141.35 & 140.61 & 141.79 \\
\hline
\end{tabular}%
\end{table}

\section{Ablation Studies}
\label{sec:ablations}


\textbf{Frame size sensitivity.} Results in  Figure~\ref{fig:frame_size_analysis} indicate that 8192 samples provide the best trade-off among the evaluated configurations. Smaller frames (2048 samples) suffer insufficient spectral resolution ($D_{\text{rel}}$ increases 18\% during transitions) and yield noisier parameter updates with audible artifacts due to controller instability, despite higher computational cost and update rates. Larger frames (16384 samples) produce smoother, more stable updates at reduced computational load, but degrade temporal tracking ($D_{\text{rel}}$ increases 12\% for 15~s transitions) due to lower update rates incompatible with typical room acoustic timescales. In the evaluated setup, the 8192-sample frame offered the most favorable compromise between spectral precision, temporal responsiveness, numerical stability, and computational feasibility. This is also the frame-size chosen by the authors in \cite{Cecchi2014}.

\vspace{6pt}

\noindent\textbf{Ground-truth LEM response.} To isolate the contribution of the online LEM estimator, we re-simulate the worst-case scenario from the main experiments — moving listener position with music excitation — replacing the live response estimate with the ground-truth LEM response for gradient computation. Figure~\ref{fig:true_LEM_ablation} summarizes validation errors for different transition lengths, showing how some optimizer instability persists even with the ground-truth response. Such effect can be attributed to the nonstationarity of the input signal. However, the instability observed around smooth acoustic transitions in the main experiments largely disappears. This confirms that most of the tested configurations are sensitive to the quality of the gradient estimate, and underlines the importance of a fast and robust online LEM estimate for stable control. 


\section{Limitations \& future work}
\label{sec:limitations}
Future work remains to fully realize the potential of the proposed framework. The present evaluation is conducted in a controlled simulation setting based on measured room impulse responses, which allows the individual components of the framework to be studied under reproducible time-varying acoustic conditions. However, this setup does not yet capture several factors that are central in practical deployments, including crowd noise at low SNR, loudspeaker and microphone nonlinearities, converter quantization, thermal drift, and uncontrolled listener movement. The reported results validate the proposed framework under the controlled linear-acoustic conditions considered here, including the specific simulation setting, plant-estimation strategy, and excitation regime. They should therefore be interpreted as evidence within that scope, rather than as a complete demonstration of live sound reinforcement performance or a universal conclusion for adaptive equalization more broadly. Future work should therefore examine robustness under realistic noise conditions, low-information frames, optimized implementations of higher-order update rules in real-world operating environments, and potential neural extensions of the LEM estimation and optimizer blocks.

\section{Conclusions}
\label{sec:conclusions}

In this work we propose a modular differentiable framework for adaptive room equalization that unifies classical adaptive filtering and DDSP-style optimization within a single closed-loop formulation. The framework was validated in simulation using measured room impulse responses under time-varying acoustic conditions, where it produced consistent equalization improvements under musical excitation. The results further suggest that frequency-domain loss functions are more appropriate than time-domain MSE for the nonstationary scenarios considered here, and that reliable online response estimation is essential for stable adaptation. Overall, the study establishes a principled link between classical Fx-LMS and differentiable audio optimization, and provides a flexible implementation for future algorithmic development and continued research on the practical deployment of adaptive room equalization. 

\section{Acknowledgments}
Funding: this work was supported by the grant FPU23/00360 (\textit{Formación de Profesorado Universitario 2023}) funded by the Spanish Ministry of Science, Innovation and Universities; and was partially funded by the Ministry of Economy and Competitiveness of Spain under grant No. PID2021-128469OB-I00.


\bibliographystyle{IEEEtranDAFx}
\bibliography{DAFx26_tmpl} 

\end{document}